# A Light-weight Distributed System for the Processing of Replicated Counter-like Objects

Joel M. Crichlow, Stephen J. Hartley

Computer Science Department, Rowan University
Glassboro, NJ, USA
crichlow@rowan.edu, hartley@elvis.rowan.edu

Michael Hosein

Computing and Information Technology Department, University of the West Indies
St. Augustine, Trinidad
mhosein2006@gmail.com

*Abstract*

*In order to increase availability in a distributed system some or all of the data items are replicated and stored at separate sites. This is an issue of key concern especially since there is such a proliferation of wireless technologies and mobile users. However, the concurrent processing of transactions at separate sites can generate inconsistencies in the stored information. We have built a distributed service that manages updates to widely deployed counter-like replicas. There are many heavy-weight distributed systems targeting large information critical applications. Our system is intentionally, relatively light-weight and useful for the somewhat reduced information critical applications. The service is built on our distributed concurrency control scheme which combines optimism and pessimism in the processing of transactions. The service allows a transaction to be processed immediately (optimistically) at any individual replica as long as the transaction satisfies a cost bound. All transactions are also processed in a concurrent pessimistic manner to ensure mutual consistency.*

## 1. Introduction

Our system is called COPAR (Combining Optimism and Pessimism in Accessing Replicas). It runs on a collection of computing nodes connected by a communications network. The transactions access data that can be fully or partially replicated. Transactions can originate at any node and the transaction processing system attempts to treat all transactions in a uniform manner through cooperation among the nodes. We have had test runs on private LANs as well as over the Internet, and preliminary results have been published and presented (see [1], [2], [3] and [4]). This paper provides some background to the project in this section, explains the basic interactions between the optimistic and pessimistic processing in section 2, and discusses recent key upgrades in sections 3 to 6.

International Journal of Distributed and Parallel Systems (IJDPS) Vol.4, No.3, May 2013One of the main reasons for replicating the data in distributed systems is to increase availability. Replication has become increasingly more useful in the face of wireless technology and roaming users. However, this replication increases the need for effective control measures to preserve some level of mutual consistency. Several replica control techniques have been proposed to deal with this issue and these techniques are described to different levels of detail in many articles and presentations (e.g. see [5], [6], [7], [8], [9], [10] and [11]).

The techniques vary in a number of ways including the number of the replicas that must participate before a change can be made to a replica, the nature of the communication among the replicas, and if a replica can be changed before the others how is the change propagated. A key contributor to the choice of specific procedures is the nature of the application. For example some applications can tolerate mutually inconsistent replicas for longer periods than others. The twin objective is

- ✓ process the transaction correctly as quickly as possible, and
- ✓ reflect this at all the replicas so that no harm is done.

One approach is to employ pessimistic strategies which take no action unless there are guarantees that consistent results and states will be generated. Such techniques sacrifice availability. Another approach is to employ optimistic techniques that take actions first and then clean up afterwards. Such techniques may sacrifice data and transaction integrity. Saito & Shapiro [12] and Yu & Vahdat [13] deal specifically with the issue of optimistic replication.

There is also the matter of failure. How do we achieve consistency and availability when the network partitions? That is when some nodes cannot communicate with other nodes. In many cases the key objective remains the same, i.e. to provide a quick response. Although that response may not be accurate it may be tolerable. Strong arguments have been made for the relaxation of consistency requirements in order to maintain good availability in the face of network partitioning. Indeed in what is referred to as Brewer's CAP theorem, the argument was made that a system can provide just two from Consistency, Availability and Partition tolerance (see [14] and [15]).

We will first demonstrate how our system works without partition tolerance to provide consistency and availability. Then we will discuss a partition tolerance implementation that maintains availability with delayed or weak consistency. Our system can be described as adhering to the Base Methodology (see [16]). That is our system conforms to the following:

- ✓ *Basically Available*: Provides a fast response even if a replica fails.
- ✓ *Soft State Service*: Optimistic processing does not generate permanent state. Pessimistic processing provides the permanent state.
- ✓ *Eventual Consistency*: Optimistic processing responds to users. Pessimistic processing validates and makes corrections.

The use of a cost bound in the processing of transactions is useful in a system where countable objects are managed. Lynch et al [17] proposed such a technique as a correctness condition in highly available replicated databases. Crichlow [18] incorporated the cost bound in a scheme that combined a simple pessimistic technique with a simple optimistic mechanism to process objects that are countable. We regard an object as countable if its data fields include only its type and how many of that object exists. For example an object may be of type blanket and there are one thousand blankets available.

Every transaction submitted to the system enters concurrently a global pessimistic two-phase commit sequence and an optimistic individual replica sequence. The optimistic sequence is moderated by a cost bound, which captures the extent of inconsistency the system will tolerate.

The pessimistic sequence serves to validate the processing and to commit the changes to the replicas or to undo an optimistic run if it generated an inconsistency. Using this scheme we built the COPAR service that can provide highly available access to counter-like replicas widely deployed over a network.

There are several examples of systems that process countable data items. Reservation systems handle available seats, rooms, vehicles, etc. Distribution systems handle available resources, e.g. blankets, bottles of water, first-aid kits and so on for disaster relief. Traffic monitoring systems count vehicles. Therefore our system COPAR although limited to countable objects has wide applicability.

The main objectives in the design were to:

- ✓ Provide a high level of availability at a known penalty to the application,
- ✓ Permit wide distribution of replicas over the network,
- ✓ Preserve data integrity, and
- ✓ Build a system that is conceptually simple.

## 2. COPAR Operation

COPAR uses the client-server model of distributed processing. Servers maintain the "database" of resources (i.e. the resource counts), and accept transactions from client machines. In our current prototype there is one client machine called the generator (it generates the transactions) and the "database" is fully replicated at all the servers. We call these servers the nodes.

Each node maintains two counts of available resources. One count is called the pessimistic or permanent count; the other count is called the optimistic or temporary count. Changes to the permanent count are synchronized with all the other nodes over the network using the two-phase update/commit algorithm (see [19], and [20]). This algorithm forces all the participating nodes to agree before any changes are made to the count. Thus, this count is the same at all the nodes and represents true resource availability. The temporary count is maintained separately and independently by each node.

In general, resource counts for availability are a single non-negative integer R, such as for one type of resource, or a vector of non-negative integers $(R_1, R_2, ..., R_m)$, such as for m types of resources. Similarly, resource counts for transactions are a single integer r, negative for an allocation and positive for a deallocation or release, or a vector of integers $(r_1, r_2, ..., r_m)$.

When the system is initialized, the permanent count $P_{jk}$ at each node j (where k ranges from 1 to m resource types) is set to the initial resource availability $R_k$. For example let $R_1$ = 2000 first aid kits, $R_2$ = 1000 blankets, or $R_3$ = 4000 bottles of water for disaster relief. Then the $P_{jk}$ for 4 nodes will be initialized as in Table 1:

Table 1. An initial state at 4 nodes

| Nodes | | | |
|---|---|---|---|
| 1 | $P_{11}$ = 2000 | $P_{12}$ = 1000 | $P_{13}$ = 4000 |
| 2 | $P_{21}$ = 2000 | $P_{22}$ = 1000 | $P_{23}$ = 4000 |
| 3 | $P_{31}$ = 2000 | $P_{32}$ = 1000 | $P_{33}$ = 4000 |
| 4 | $P_{41}$ = 2000 | $P_{42}$ = 1000 | $P_{43}$ = 4000 |

The temporary count $T_{jk}$ at each node is set to the initial permanent count divided by the number of nodes n. $T_{jk}$ is then adjusted upward by an over-allocation allowance c, called the cost bound, where c >= 1. Therefore,

$T_{jk} = c * P_{jk} / n$

For example, if there are four nodes, if $R_1$ is 100, and if c is 1.16, then $Pj_1$ is set to 100 and $T_{j1}$ is set to 29 at each node as in Table 2:

Table 2. Initial permanent and temporary counts at 4 nodes

| Nodes | |
|---|---|
| 1 | $P_{11} = 100$ <br> $T_{11} = 29$ |
| 2 | $P_{21} = 100$ <br> $T_{21} = 29$ |
| 3 | $P_{31} = 100$ <br> $T_{31} = 29$ |
| 4 | $P_{41} = 100$ <br> $T_{41} = 29$ |

Most reservation/allocation systems allow some over-allocation to compensate for reservations that are not used, such as passengers not showing up for an airline flight or people not picking up supplies when delivered to a relief center. There is a cost involved in over-allocation, such as compensating passengers denied boarding on an airline flight. Organizations using a reservation/allocation system must carefully evaluate the cost of over-allocation and limit it to what can be afforded or tolerated.

Currently, interaction with the system is simulated by generating a transaction $t_i$, which makes a request ($r_1, r_2, ..., r_m$), i.e. for $r_i$ resources of type i, where i ranges from 1 to m types of resources. This request is sent to a node j. This node is then considered the parent or owner of the transaction.

The m integers in a transaction are generated randomly and the node j is chosen at random from 1 to n, where there are n nodes. Transactions from the generator are numbered sequentially. Additions to the pool of resources are handled differently. Such additions are discussed in section 4.

For example, a transaction may make a request (-10, -20, -100) for 10 first aid kits, 20 blankets and 100 bottles of water, where there are 3 resource types: type1 – first aid kits, type 2 – blankets and type 3 – bottles of water. Furthermore a transaction deallocating or returning 10 first aid kits, 20 blankets and 100 bottles of water may be expressed as (10, 20, 100).

Each node maintains two queues of transactions, called the parent or owner queue and the child queue. A parent node, on receiving a transaction, adds that transaction to its parent queue and to its child queue. The transaction is also broadcast to all nodes to be added to each node's child queue. The transactions $t_i$ in each node's parent queue, are kept sorted in order of increasing i, in other words, in the order generated by the transaction generator.

Note that a particular transaction $t_i$ is in exactly one node's parent queue. Each node j has two processors (threads), one responsible for maintaining the parent queue and the permanent count $P_{jk}$ at the node, and the other responsible for maintaining the child queue and the temporary count $T_{jk}$ at the node (see Figure 1).

The permanent processor at each node participates in a two-phase commit cycle with all the other node permanent processors. After the processing of transaction $t_{i-1}$ by its parent, the node whose parent queue contains transaction $t_i$ becomes the coordinator for the two-phase commit cycle that changes the permanent count $P_{jk}$ at all nodes j to $P_{jk} + r_k$ for k = 1, 2, ...,m. The temporary counts are also forced to change after permanent processing. We will discuss this in the following section.

The change to the permanent count is subject to the restriction that $P_{jk} + r_k$ is nonnegative for all k. If that is not the case, all $P_{jk}$ are left unchanged and the transaction $t_i$ is marked as a violation. This in effect means that if a request cannot be totally granted then nothing is granted. (This will be upgraded during further research to allow non-violation if $P_{jk} + r_k$ is nonnegative for at least one k, i.e. to allow granting of partial requests). At the end of the two-phase commit cycle, the owner (parent) of transaction $t_i$ sends a message to all nodes, including itself, to remove $t_i$ from the node's child queue if $t_i$ is present.

Temporary processing takes place concurrently with permanent processing. The temporary processor at each node j removes the transaction $t_h$ at the head of its child queue, if any, and calculates if the request $r_k$ made by $t_h$ can be allocated or satisfied from its temporary (optimistic) count $T_{jk}$. In other words, node j checks if it is the case that $T_{jk} + r_k$ is non-negative for all k = 1, 2, … m. If that is not the case, $t_h$ is discarded (This will be upgraded during further research so that transaction $t_h$ is not discarded if $T_{jk} + r_k$ is nonnegative for at least one k); otherwise, node j sets $T_{jk}$ to $T_{jk} + r_k$ and sends a message to the parent (owner) node of the transaction, i.e. the node whose parent queue contains the transaction.

When a parent node n receives such a message from node j for transaction $t_h$, node n makes two checks.

• Is this the first such message received from any node's temporary processor for transaction $t_h$?
• Has transaction $t_h$ been done permanently yet?

If this is not the first such message, a reply is sent to node j that it should back out of the temporary allocation it did for $t_h$, that is, change its temporary count $T_{jk}$ to $T_{jk} − r_k$. This operation is necessary since another node will have done the temporary processing. This is possible because all the nodes get a chance to make an individual response to a request. The fastest one wins.
A temporary transaction may have to be "undone". Therefore, if this is the first such message and if the transaction $t_h$ has not yet been done permanently (pessimistically), node j sending the message is marked as the node having done transaction $t_h$ temporarily (optimistically). If this is the first such message, but transaction $t_h$ has already been done permanently, no node is recorded as having done the transaction temporarily.

When the permanent processor in a node j coordinates the two-phase commit for a transaction $t_i$ and has decided that transaction $t_i$ is a violation, that is, $P_{jk} + r_k$ is negative for one or more k, node j checks to see if the transaction was marked as having been done optimistically earlier by some node's temporary processor. If so, the transaction $t_i$ is marked as "undone," meaning that a promised request cannot now be granted.

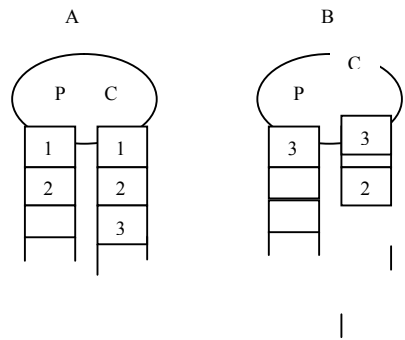

Figure 1. Each node A and B has a parent queue P and a child queue C. Node A owns transactions 1 and 2 and will process these pessimistically in a two-phase commit protocol involving A and B. Node B owns transaction 3 and will process it pessimistically in a two-phase commit protocol involving A and B. Concurrently nodes A and B process transactions 1, 2 and 3 optimistically.

If no node has done the transaction optimistically and it is not a violation, the owner's temporary processor allocation $T_{jk}$ is "charged" for it, $T_{jk} = T_{jk} + r_k$. This is done to lessen the probability of a later transaction being performed optimistically but then marked "undone" by the permanent processor.

## 3. Updating Optimistic counts after Pessimistic/Permanent Processing

The temporary optimistic counts $T_{jk}$ change at the end of optimistic transaction processing. The pessimistic counts $P_{jk}$ change at the end of permanent transaction processing. Whenever there is a change to $P_{jk}$ this should generate an update to $T_{jk}$ which is consistent with the new $P_{jk}$. Therefore $T_{jk}$ is updated by the temporary optimistic processing and after the pessimistic permanent processing.

As is stated above, when the system is initialized, the permanent count $P_{jk}$ at each node j (where k ranges from 1 to m resource types) is set to the initial resource availability $R_k$. However, permanent processing of a transaction generates a new $P_{jk}$ where

(new) $P_{jk}$ = (old) $P_{jk} + r_k$

Therefore a new $T_{jk}$ is generated where

$T_{jk} = c * $ (new) $P_{jk} * w_{jk}$

You may notice that there is an apparent difference between how $T_{jk}$ is derived here and how it was derived initially (/n is replaced by $* w_{jk}$). The $w_{jk}$ is a weight which captures the amount of allocations done by a node and influences the reallocation of the $T_{jk}$ values.

We will now discuss how the $w_{jk}$ is calculated. Permanent processing uses the two-phase commit protocol which requires a coordinating node. Permanent processors via the coordinator maintain a running total of allocations done by each node. Let $ra_{jk}$ be the total allocations of resource k done by node j on completion of permanent processing. Let $RA_k$ will be the total allocations of resource k done by all the nodes at end of permanent processing. We let

$w_{jk} = (ra_{jk} + 1)/(RA_k + n)$ where n is the number of nodes.

Note that initially $ra_{jk}$ and $RA_k$ are equal to zero, therefore on initialization $w_{jk}$ is equal to 1/n. This is consistent with the initial derivation of $T_{jk}$.

The coordinating parent processor can now use

$T_{jk} = c * P_{jk} * w_{jk}$

to compute the new temporary counts for optimistic processing. However there is a problem here. While the coordinating pessimistic processing was being done, the optimistic temporary processors were still running. Therefore the information used in the computation of the $T_{jk}$ can be stale. That is the $ra_{jk}$ used in the computation of the new $T_{jk}$ for node j could have been changed due to further optimistic processing by that node.

We must therefore distinguish between two $ra_{jk}$. Let the one that was used by pessimistic processing to compute the new count be called $ra_{jk,recorded}$ and the current one be $ra_{jk,current}$.

When the temporary processors receive $T_{jk}$ from the permanent processor the temporary processors adjust the $T_{jk}$ as follows in order to reflect its current position:

$T_{jk} = T_{jk} - (ra_{jk,current} - ra_{jk,recorded})$

If this result is negative make it 0. The zero value forces temporary processing at this node to stop.

For example, let $R_1$ denote resources of type 1 (say blankets) initially 100 and $c = 1.1$.

Then $P_{j1} = 100$ and $T_{j1} = 110/n$.
Let there be 3 replicas i.e. $n = 3$.
Therefore $T_{11} = T_{21} = T_{31} = 37$.

Let permanent processors at nodes 1, 2, 3 record allocations of 30, 20, 10 blankets respectively.

Therefore
$ra_{11,recorded} = 30$, $ra_{21,recorded} = 20$, and $ra_{31,recorded} = 10$.
Therefore $P_{j1} = R_1$ is now 40 (i.e. $100 - 30 - 20 - 10$) and
$T_{j1} = 1.1 * 40 * w_{j1}$

Therefore
$T_{11} = ((30 + 1) / (60 + 3)) * 44 = 22$

Assume that 6 more blankets were allocated at temporary processor 1.

Therefore
$T_{11} = T_{11} - (ra_{11,current} - ra_{11,recorded}) = 22 - (36 - 30) = 16$

We now compute the new temporary count for temporary processor 2.
$T_{21} = ((20 + 1) / (60 + 3)) * 44 = 15$

Assume 4 more blankets were allocated at temporary processor 2.

Therefore
$T_{21} = T_{21} - (ra_{21,current} - ra_{21,recorded}) = 15 - (24 - 20) = 11$

We now compute the new temporary count for temporary processor 3.
$T_{31} = ((10 + 1) / (60 + 3)) * 44 = 7$

Assume 3 blankets were returned/deallocated at temporary processor 3.

Therefore
$T_{31} = T_{31} - (ra_{31,current} - ra_{31,recorded}) = 7 - (7 - 10) = 10$

On the other hand let's assume that 8 more blankets were allocated at temporary processor 3, then $ra_{31,current} = 18$, and

$T_{31} = T_{31} - (ra_{31,current} - ra_{31,recorded}) = 7 - (18 - 10) = -1$. This temporary count is then set to 0 and temporary processor 3 is stopped until it gets a count greater than 0.

International Journal of Distributed and Parallel Systems (IJDPS) Vol.4, No.3, May 2013Note that this still does not prevent temporary over-allocations since one temporary does not know what the other temporary is doing and cost bound c = 1.1. However, it reduces the incidents of over-allocations and hence the number of "undones". But our objective of high-availability is being maintained.

## 4. ADDITIONS

At any time while the system is running additions can be made to the available pool of resources, e.g. new donations can be made to a pool of resources for disaster relief. An addition is considered a unique transaction called $a_i(r_1 \ldots r_m)$ that adds $r_k$, (i.e. r resources of type k where k ranges from 1 to m) to the pool of available resources. It is not appended to the child queues. When this transaction is processed $P_{jk}$ and $T_{jk}$ are updated by the permanent processor:

$P_{jk} = P_{jk} + r_k$

$T_{jk} = c * P_{jk} * w_{jk}$ using the current values of the $w_{jk}$.

The temporary processors will then update $T_{jk}$ to reflect their current situation as discussed in section 3.

## 5. RESULTS FROM TESTS WHERE THERE ARE NO FAILURES

The COPAR test-bed includes a transaction generator and servers on a LAN at Rowan University in New Jersey (R) interconnected via the Internet with a server about 40 miles away at Richard Stockton College in New Jersey (RS) and a server at the University of the West Indies (UWI) located in Trinidad in the southern Caribbean approximately 2000 miles away.

The transaction generator and servers are all started with a program running on the transaction generator node that reads and parses an XML file containing the data for the run. We have demonstrated that a large percentage of transactions submitted to the system can be handled optimistically (without multi-server agreement) at significantly faster turnaround times and with a very small percentage of "undones".

The figures and tables display results when 200 transactions were generated at a rate of 5 transactions per second. There are 200 resources of each type available initially. The cost bound is 1.16. Transactions include requests for resources, returns of resources and new donations (i.e. additions). Requests and returns range from 3 to 9 resources per transaction. Donations range from 3 to 10 resources per donation. Tests were done on two platforms: a four node platform with all servers (including transaction generator) at Rowan (R); and a six node platform including Rowan, Richard Stockton and UWI (R+RS+UWI).

In Figure 2, of the 200 transactions 18 are donations totaling 136. There is one resource type of 200 resources available initially. On the R platform 159 transactions are done optimistically and 4 are undone. On the R+RS+UWI platform 182 transactions are done optimistically and 25 are undone.

During these tests on the R platform, pessimistic processing times (PT) range from 29 milliseconds to 288 milliseconds, optimistic processing times (OT) range from 1 millisecond to 20 milliseconds. The average PT to RT ratio is 18. The R+RS+UWI platform is subject to the vagaries of the Internet and the vast geographical expanse. The PT times range from 2.6 seconds to 10 minutes, OT times range from 1 millisecond to 1 second. The average PT to RT ratio is 117000 (see Table 3).

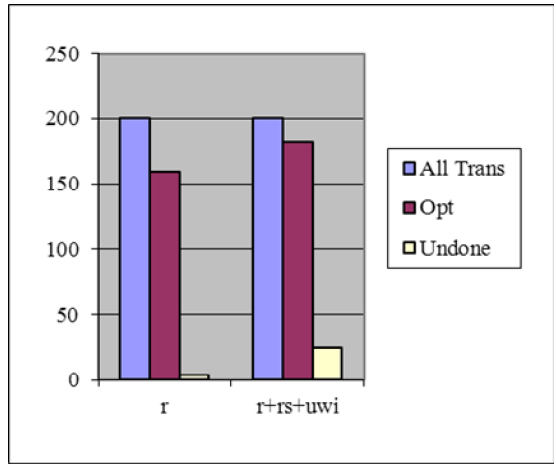

Figure 2. Tests were done on two platforms: a four node platform with all servers (including transaction generator) at Rowan (R); and a six node platform including Rowan, Richard Stockton and UWI (R+RS+UWI); 200 transactions were generated at a rate of 5 transactions per second.

Table 3. Pessimistic (PT) and Optimistic (OT) times from tests in Figure 2

|  | Min | Max | Ave PT/OT ratio |
|---|---|---|---|
| R PT | 29 msec | 288 msec | 18 |
| R OT | 1 msec | 20 msec |  |
| R+RS+UWI PT | 2.6 sec | 10 min | 117000 |
| R+RS+UWI OT | 1 msec | 1 sec |  |

In Figure 3 there are 3 resource types each with 200 resources initially. There are 28 donations totaling 189, 172 and 181 for resource types 1, 2 and 3 respectively. On the R platform 161 transactions are done optimistically and 2 are undone. On the R+RS+UWI platform 172 transactions are done optimistically and 10 are undone.

During these tests on the R platform, PT times range from 29 milliseconds to 255 milliseconds, OT times range from 1 millisecond to 21 milliseconds. The average PT to RT ratio is 19. The R+RS+UWI platform is subject to the vagaries of the Internet and the vast geographical expanse. The PT times range from 3 seconds to 8 minutes, OT times range from 1 millisecond to 258 milliseconds. The average PT to RT ratio is 111000, (see Table 4).

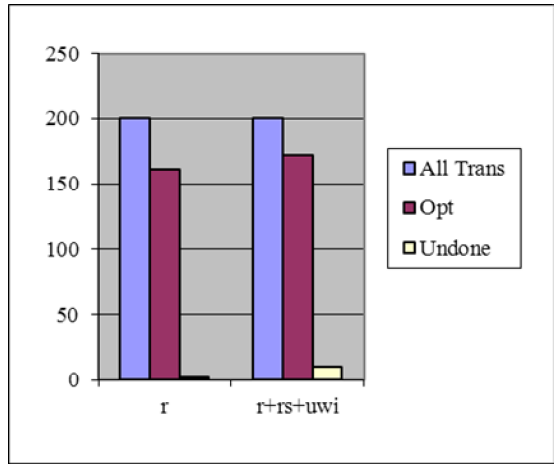

Figure 3. There are 3 resource types each with 200 resources initially. There are 28 donations totaling 189, 172 and 181 for resource types 1, 2 and 3 respectively. On the R platform 161 transactions are done optimistically and 2 are undone. On the R+RS+UWI platform 172 transactions are done optimistically and 10 are undone

Table 4. Pessimistic (PT) and Optimistic (OT) times from tests in Fig. 3

|  | Min | Max | Ave PT/OT ratio |
|---|---|---|---|
| R PT | 29 msec | 255 msec | 19 |
| R OT | 1 msec | 21 msec |  |
| R+RS+UWI PT | 3 sec | 8 min | 111000 |
| R+RS+UWI OT | 1 msec | 258 msec |  |

## 6. HANDLING FAILURE

Our failure handling model addresses only the case of a node that can no longer be reached. Failing to reach a node may be due to that node's failure, communication link failure, or an unacceptably long response time. Such a failure handling model is workable in COPAR since the transactions handled and the information maintained by the system can tolerate certain margins of error.

If a node cannot be reached due to node or communication link failure then the pessimistic 2PC processing will fail. However optimistic processing will continue at all operating nodes until the cost bound at those nodes is zero. The objective will be to restart pessimistic processing only if a majority of the initial group of nodes can be reached.

The restart of pessimistic processing among the majority uses the concept of the "distinguished partition". That is, the network is now partitioned and processing is allowed to continue in a favored partition. This favored partition is called the "distinguished partition". Voting schemes in which nodes possess read/write votes are often used to determine that "distinguished partition" (see [21] and [9]).

International Journal of Distributed and Parallel Systems (IJDPS) Vol.4, No.3, May 2013

Our "distinguished partition" for pessimistic processing will be the partition with the majority of the initial group of nodes. The restart will use the current permanent/pessimistic resource counts and generate new temporary/optimistic counts for the new reachable pool of nodes.

For example, given the following 4-node situation in Table 5:

Table 5. The current state at 4 nodes

| Nodes |  |
|---|---|
| 1 | $P_{11} = 100$<br>$T_{11} = 29$ |
| 2 | $P_{21} = 100$<br>$T_{21} = 29$ |
| 3 | $P_{31} = 100$<br>$T_{31} = 29$ |
| 4 | $P_{41} = 100$<br>$T_{41} = 29$ |

After some processing, assume that each node has allocated 4 resources and this has been processed pessimistically, therefore the new situation is Table 6:

Table 6. The new state after allocating 4 resources

| Nodes |  |
|---|---|
| 1 | $P_{11} = 84$<br>$T_{11} = 25$ |
| 2 | $P_{21} = 84$<br>$T_{21} = 25$ |
| 3 | $P_{31} = 84$<br>$T_{31} = 25$ |
| 4 | $P_{41} = 84$<br>$T_{41} = 25$ |

Assume that node 4 can no longer be reached, but it is still operable. That node can continue to process requests until $T_{41} = 0$. Pessimistic processing can restart with nodes 1, 2 and 3 with a P value of 84 and 3 being the number of nodes.

Currently the system is controlled by a transaction generator, which can be viewed as a single point of failure. In the future transaction handling should be separated from system management. The system manager will handle system start-up, initialization, monitoring, restart, etc. The transaction handlers should exist at all nodes. In the meantime the transaction generator assumes the role of the monitor of the system.

We would like the two-phase commit processing to recover from the failure of a participating node. Therefore we are proposing the following pseudo two-phase commit sequence.

In phase one, after a time-out before receiving all votes, the coordinator will count the number of votes to determine if it heard from a majority of the initial set of participants. If it did not hear from a majority the transaction will be aborted. If it heard from a majority it will start phase two with the majority as the group of participants.

In phase two, after a time-out before receiving all commit responses, the coordinator will determine if it heard from a majority of the initial set of participants. If it did not hear from a majority the transaction will be aborted. If it heard from a majority the coordinator will complete the commit sequence. The subsequent processing round will start phase one with this new group of participants.

If the transaction generator times out on a coordinator it will assume that the coordinator is no longer reachable. The transaction generator will determine if a majority of the initial set of nodes is still operable. If a majority is operable the transaction generator will select a new coordinator and restart transaction processing with the new group of nodes. If a majority is not operable the transaction generator will wait until a majority of nodes come back online.

As a proof of concept we ran some tests of COPAR that simulated a node failure during transaction processing. We did this in the following way. Whenever a server is notified that a transaction has been selected for pessimistic processing, that server increments a counter that was initially zero. When that counter reaches 50 the server checks the notification message to determine if the sender was a specified server s. If it is s then s is classified as inactive and is dropped from the two-phase commit pool. The pessimistic processing continues with the pool reduced by one server.

However, since server s is in reality still active it will continue optimistic processing until it empties its child queue or until its cost bound is less than or equal to zero. At this point the transaction generator does not know that server s is no longer in the two-phase commit pool and so the generator can continue to send new transactions to server s.

In order to prevent this, the generator increments a counter whenever it generates a new transaction. When that counter reaches 25 the generator stops sending transactions to server s. Transactions that should have gone to s are sent alternately to its downstream and upstream neighbor. In the tests 200 transactions are generated. Therefore the counter values must be less than 200. Since the selection of a coordinator/parent is pseudo-random and since we do not keep a history of the interactions between servers then our choice of the counter values are somewhat arbitrary, and it is intended primarily to ensure that new transactions are not sent to server s after it has been dropped from the pessimistic two-phase pool.

In the tests discussed below a server on the Rowan(R) LAN is dropped during the processing. In Figure 4, of the 200 transactions 18 are donations totaling 136. There is one resource type of 200 resources available initially. On the R platform 157 transactions are done optimistically and 1 is undone. On the R+RS+UWI platform 182 transactions are done optimistically and 25 are undone. Notice the similarity between these results and those displayed in Figure 2 where the only change here is in the dropped server.

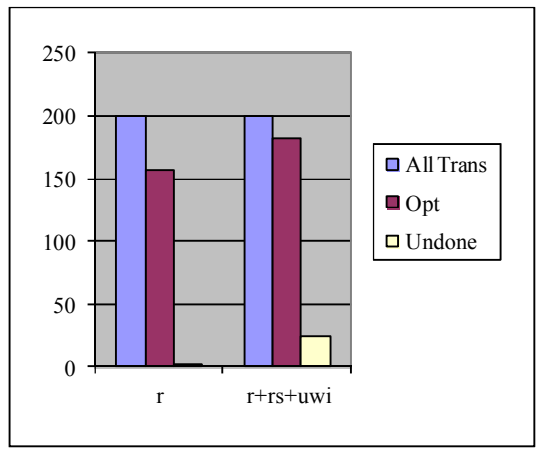

Figure 4. A server on the Rowan(R) LAN is dropped during the processing. Results are similar to case when no server is dropped.

During these tests on the R platform, pessimistic processing times (PT) range from 29 milliseconds to 222 milliseconds, optimistic processing times (OT) range from 1 millisecond to 17 milliseconds. The average PT to RT ratio is 21. The R+RS+UWI platform is subject to the vagaries of the Internet and the vast geographical expanse. The PT times range from 2.6 seconds to 8.8 minutes, OT times range from 1 millisecond to 991 milliseconds. The average PT to RT ratio is 117000 (see Table 7).

Notice that whereas the numbers of completions are similar to the case when all servers were operable (see Table 3), there are differences in completion times when a server is dropped. It is expected that the pessimistic processing should decrease after a server was dropped. On the R platform max PT dropped from 288 milliseconds to 222 milliseconds, and on the R+RS+UWI platform max PT dropped from 10 minutes to 8.8 minutes.

Table 7. Pessimistic (PT) and Optimistic (OT) times from tests in Fig. 4

|  | Min | Max | Ave PT/OT ratio |
|---|---|---|---|
| R PT | 29 msec | 222 msec | 21 |
| R OT | 1 msec | 17 msec |  |
| R+RS+UWI PT | 2.6 sec | 8.8 min | 97251 |
| R+RS+UWI OT | 1 msec | 991 msec |  |

In Figure 5, of the 100 transactions 9 are donations totaling 66. There is one resource type of 200 resources available initially. The results of two tests on the R+RS+UWI platform are displayed. In both tests a Rowan server was dropped after about 50 transactions. In the test labeled "more" the distribution of transactions was such that the remote UWI server (about 2000 miles away from the generator) got 50% more transactions than in the test labeled "less". In each case 91 transactions are done optimistically and 0 is undone. The difference in the distribution of transactions does not affect the numbers completed.

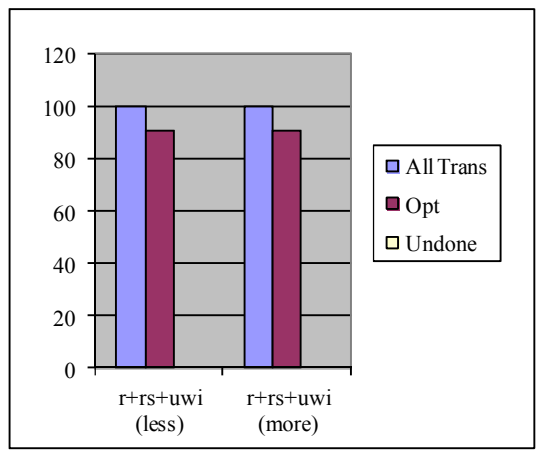

Figure 5. The results of two tests on the R+RS+UWI platform are displayed. In both tests a Rowan server was dropped after about 50 transactions. In the test labeled "more" the distribution of transactions was such that the remote UWI server got 50% more transactions than in the test labeled "less". In each case 91 transactions are done optimistically and 0 is undone.

During these tests on the "less" platform, pessimistic processing times (PT) range from 2.8 seconds to 4 minutes, optimistic processing times (OT) range from 1 millisecond to 835 milliseconds. The average PT to RT ratio is 47389. On the "more" platform the PT times range from 4.4 seconds to 6 minutes, OT times range from 2 millisecond to 1.4 seconds. The average PT to RT ratio is 45674 (see Table 8). On the "more" platform the far-distant UWI server performed the coordinator role more often than on the "less" platform. Therefore the nature of the two-phase commit would generate a longer max PT time. However to the satisfaction of the users of the system the maximum optimistic processing time is 1.4 seconds with 0 undone.

Table 8. Pessimistic (PT) and Optimistic (OT) times from tests in Fig. 5

|         | Min    | Max      | Ave PT/OT ratio |
|---------|--------|----------|-----------------|
| Less PT | 2.8 sec | 4 min   | 47389           |
| Less OT | 1 msec | 835 msec |                 |
| More PT | 4.4 sec | 6 min   | 45674           |
| More OT | 2 msec | 1.4 sec  |                 |

## 7. CONCLUSION

We feel that we have met the main objectives that we had set for COPAR. It targets applications where there is need for very fast receipt and distribution of resources over possibly wide geographical areas, e.g. a very wide disaster zone. COPAR provides a high level of availability. There is very fast turnaround time on the processing of transactions. The validation is quick thus minimizing the need to undo an optimistic result. There is a simple failure handling scheme which permits all reachable nodes to continue optimistic processing and a "distinguished partition" to continue pessimistic processing.

There is wide geographical distribution of replicas covering a range of approximately 2000 miles. Data integrity is preserved through the pessimistic two-phase commit and the choice of an initial cost bound. It is our view that the design embodies simple but workable concepts. All nodes handle their child queues optimistically (independently) and their parent queues pessimistically (two-phase commit).

However there is further work to be done. Three main tasks are (1) improving the handling of failure, (2) separating the system manager from the transaction manager and (3) implementing multiple transaction generators with interfaces that run on mobile devices.

International Journal of Distributed and Parallel Systems (IJDPS) Vol.4, No.3, May 2013